\documentclass[final,5p,times,twocolumn]{elsarticle}

\usepackage{graphicx}
\usepackage{amsmath}
\usepackage{amssymb}
\usepackage{dcolumn}
\usepackage{endnotes}

\usepackage[T1]{fontenc}
\usepackage[sc]{mathpazo}
\usepackage[usenames]{color}

\newcommand{\Fe}{\text{Fe}}
\newcommand{\Cr}{\text{Cr}}
\newcommand{\He}{\text{He}}
\newcommand{\eV}{\text{eV}}
\newcommand{\meV}{\text{meV}}
\newcommand{\MeV}{\text{MeV}}

\newcommand{\fref}[1]{Fig.~\ref{#1}}
\newcommand{\Fref}[1]{Figure~\ref{#1}}
\newcommand{\tref}[1]{Tab.~\ref{#1}}

\newcommand{\sref}[1]{Sec.~\ref{#1}}

\usepackage{ulem}

\renewcommand{\emph}[1]{{\it #1}}

\journal{Journal of Nuclear Materials}

\begin{document}

\begin{frontmatter}



\title{Calculation of the substitutional fraction of ion-implanted He in an Fe target}

\author{Paul Erhart and Jaime Marian}

\address{Lawrence Livermore National Laboratory, Livermore, CA 94551}

\begin{abstract}
Ion-implantation is a useful technique to study irradiation damage in nuclear materials. To study He effects in nuclear fusion conditions, He is co-implanted with damage ions to reproduce the correct He/dpa ratios in the desired or available depth range. However, the short-term fate of these He ions, \emph{i.e.} over the time scales of their own collisional phase, has not been yet unequivocally established. Here we present an atomistic study of the short-term evolution of He implantation in an Fe substrate to approximate the conditions encountered in dual ion-implantation studies in ferritic materials. Specifically, we calculate the fraction of He atoms that end up in substitutional sites shortly after implantation, \emph{i.e.} before they contribute to long-term miscrostructural evolution. We find that fractions of at most 3\%\ should be expected for most implantation studies. In addition, we carry out an exhaustive calculation of interstitial He migration energy barriers in the vicinity of matrix vacancies and find that they vary from approximately 20 to 60 meV depending on the separation and orientation of the He-vacancy pair.
\end{abstract}


\end{frontmatter}


\section{Introduction}
\label{sec:intro}

This paper tries to answer a simple yet important question in He-implantation studies: Do ion-implanted He atoms end up as interstitial or as substitutional particles in the target matrix? The difference is critical because of the large diffusivity difference between both forms of He: interstitial He (i-He) diffuses extremely fast, sampling large portions of the configurational space quickly, readily finding other defects or microstructural features. Conversely, substitutional He (s-He), while energetically more stable, is immobile, necessitating migration of other point defects before it can move. This can happen either by reacting with a self-interstitial atom (SIA) that recombines with the vacancy and knocks the He back to an interstitial site, or by correlated lattice exchange reactions with a vacancy in nearest-neighbor positions. Either way, the diffusivity of s-He is still several orders of magnitude lower than that of i-He.

Despite the important implications of these mechanisms on the subsequent microstructural evolution, at present most researchers consider that 100\%\ of the implanted He is interstitial \cite{yoo1979, ghoniem1983, russell2008} and can only become substitutional by finding an isolated vacancy in an uncorrelated fashion via long-range diffusion. The question we ask here is whether this is true for all He atoms or whether some of them can become substitutional as part of their own implantation process prior to uncorrelated diffusion taking place. 

Here we present a computational study involving the binary collision approximation (BCA), molecular dynamics (MD), and kinetic Monte Carlo (kMC) simulations. The BCA is used to simulate the penetration of He beams of various energies into Fe targets, and to obtain energy distributions of Fe recoils due to He impact. MD is then used to simulate He thermalization after the primary knock-on event in its own collisional environment, and to ascertain whether He atoms create stable Frenkel pairs that can result in correlated recombination. Finally, we use kMC to calculate the fraction of \emph{freely-migrating} He atoms from those that do create defects but do not find the vacancy during MD time scales. From these simulations, we find that nearly 3\%\ of the He atoms end up in subsitutional sites. While this number appears small, it nonetheless leads to dramatic differences in the microstructural evolution of the material, as will be shown in future studies.

\section{Results}

\subsection{Calculation of recoil distributions and He energies}
\label{subsec:srim}

\begin{figure}
  \centering
  \includegraphics[width=0.9\columnwidth]{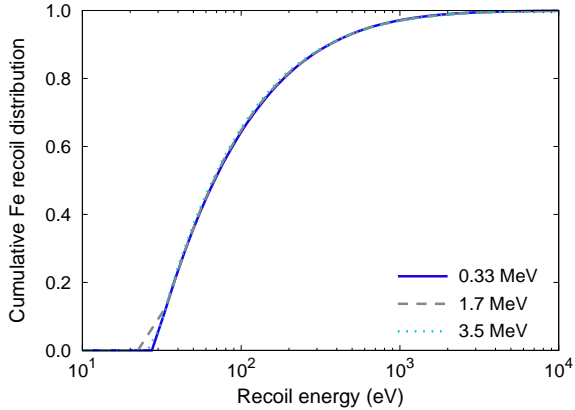}  
  \caption{
    Cumulative Fe recoil distribution for He-ion beam irradiations with recoil energies of 0.33, 1.7 and 3.5 MeV incident energy. The threshold displacement energy is 25 eV.
  }
  \label{fig:srim}
\end{figure}

\begin{table}
  \centering
  \caption{
    SRIM parameters for the three He-beam energies considered.
  }
  \label{tab:t_ions}
  \begin{tabular}{lccc}
    \hline\hline
    He ion energy (MeV)         &   0.33 &   1.7  &   3.5  \\
    \hline
    Depth ($\mu$m)              &   0.7  &   2.6  &   6.0  \\
    \%\ energy to recoils       &   0.43 &   0.11 &   0.07 \\
    Average recoil energy (eV)  & 194    & 211    & 222    \\
    Maximum recoil energy (keV) &  66    & 134    & 315    \\
    \hline\hline
  \end{tabular}
\end{table}

The He energy range of interest for fusion materials lies between 3.5 and 0.33\,MeV, corresponding to the energy of $\alpha$ particles emitted from fusion reactions and those produced via (n,$\alpha$) transmutation reactions in Fe\endnote{The transmutation reaction $^{56}$Fe(n,$\alpha$)$^{53}$Cr results in an excess mass of $m_e=m_{\Fe}+m_{\text{n}}-(m_\Cr+m_\alpha)=55.9349375+1.0086692-(52.9406494+4.0026032)=0.0003541\,\text{amu}$. This is equivalent to $E=m_ec^2=0.33\,\MeV$, although some variability in the form of a relatively narrow energy spectrum is to be expected depending on local conditions such as orientation, atomic vicinity, etc. (Source of particle rest masses: NIST \cite{nist})}. In addition, ion beam experiments typically use energies in this range to achieve penetrations of a few microns, so it is useful to have an intermediate energy for reference. Thus, we first calculate the Fe recoil distribution for three He-ion energies, namely 0.33, 1.7 and 3.5\,MeV, using SRIM \cite{ziegler1984}. The cumulative recoil energy distributions in each case are given in \fref{fig:srim}, where a threshold displacement energy of 25\,eV was used. As the figure shows, the three recoil distributions are almost identical. This is because He ions only create recoils when they have slowed down to a few keV, without much participation from their higher-energy histories, which as shown in \tref{tab:t_ions} only contribute to penetration and maximum recoil energy. It is more informative to compare the average recoil energies, which, in contrast, differ only by a few eV. In all cases, the energy expended in recoils amounts to less than 0.5\%\ of the total ion energy.

\begin{figure}
  \centering
  \includegraphics[width=0.9\columnwidth]{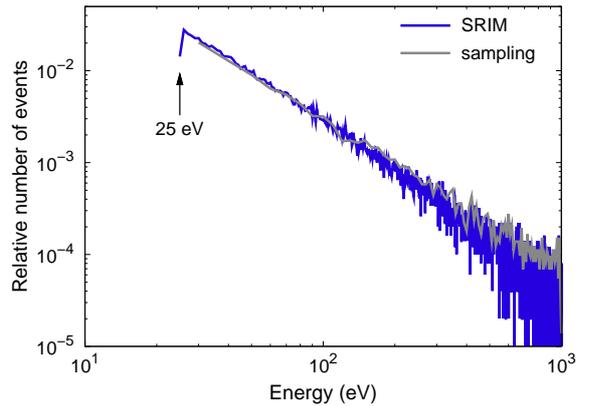}  
  \caption{
    Spectrum of Fe recoils sampled by MD simulations in comparison with spectrum for 1.7\,MeV He obtained using SRIM.
  }
  \label{fig:sampling}
\end{figure}

These results suggest that in the energy range relevant to fusion materials the actual ion energy is irrelevant for damage purposes. Therefore, we take the 1.7\,MeV spectrum shown in \fref{fig:sampling} as representative of all He energies and proceed to simulate the effect of these recoils on lattice damage. We note that SRIM does not capture channeling, which may have some impact on the final results.

\subsection{Molecular dynamics simulations of He impact in Fe}
\label{subsec:md}

Next we study the fate of He ions at the end of their collision trajectories, when they collide with the last of their recoils before thermalizing in the host lattice. We assume that these recoils are ejected with energies consistent with the recoil spectrum obtained using SRIM (see \fref{fig:sampling}), as the probability for high energy recoils, produced early in the He collision sequences, is very small. The objective is then to investigate whether He ions can become substitutional by interacting with vacancies of their own creation, rather than by long-range diffusion. For this, we perform MD simulations of He collisions in Fe and analyze the final configurations.

Simulations were carried out using the massively parallel MD code \textsc{lammps} \cite{lammps}. The simulation cell consisted of $38\times38\times38$ conventional body-centered cubic (BCC) unit cells corresponding to 109,744 Fe atoms. All simulations were carried out at a temperature of 700\,K, which is representative of fusion conditions. Atomic interactions were modeled using the Fe-He interatomic potential of developed by Juslin and Nordlund \cite{juslin2008}, which builds on the Fe potential by Mendelev {\it et al.} \cite{MenHanSro03} and gives an equilibrium lattice constant of $a_0=2.871\,\text{\AA}$ at 700\,K.
The collision simulations were done with He as an energetic ion in an otherwise perfect Fe lattice. The He atoms was assigned kinetic energies that produce a Fe recoil distribution consistent with the SRIM data in \fref{fig:sampling}. We sample the Fe recoil velocity $v_{\Fe}$ directly from the SRIM spectrum and, assuming purely elastic collisions, assign a velocity $v_\He$ to the He atom 
\begin{equation}
  v_\He = \left(\frac{m_\He}{m_\Fe}-1\right) v_\Fe.
  \label{eq:v}
\end{equation}
where the mass ratio is $m_{\mathrm{He}}/m_{\Fe}$$\approx 0.077$ and the velocity vector is directed at the Fe PKA.

\begin{figure}
  \centering
  \includegraphics[width=0.9\columnwidth]{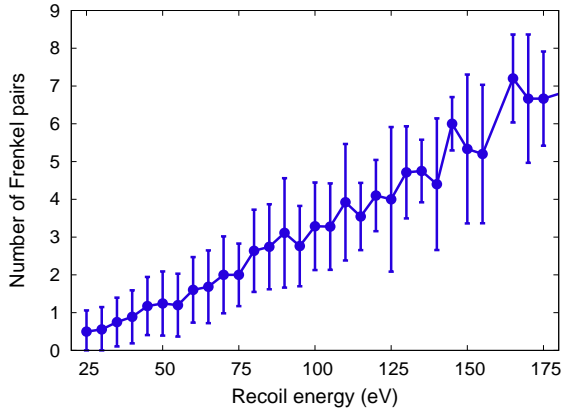}  
  \caption{
    Average number of stable Frenkel pairs created during the MD simulations of He collision cascades in Fe. The bars indicate the standard deviation of the number of defects created.
  }
  \label{fig:fp}
\end{figure} 

The atoms that were initially within a radius of $18.65\,a_0$ of the center of the simulation cell were considered the core region while all other atoms were assigned to the edge region. The latter were coupled to a Nos\'e-Hoover thermostat at 700\,K, while the remainder of the system evolved in time according to the microcanonical ensamble. The equations of motion were integrated until either the He atom thermalized (kinetic energy less than 0.1\,eV) or escaped the core region. The final configuration was relaxed using conjugate gradient minimization and defects were identified by analyzing the occupancies of the Wigner-Seitz cells of the initial perfect BCC lattice. In total, we simulated 5,000 events to extract sufficient statistics.

Out of the total 5,000 events simulated, 2,668 (53.4\%) were terminated when the He atom exited the core region. They are interpreted as cases in which the He equilibrates in a region ``far'' away from the PKA, such that the probability for the He to recombine with defects created in the original cascade is very small. Accordingly these events are counted as i-He in the total atom tally. The remaining 2,332 events were analyzed to obtain the number and distribution of irradation induced point defects. In 25 cases the helium ended up in a substitutional site corresponding to 1.07\%\ of the ``thermalized'' cases and 0.5\%\ of the total number of events. This occurred typically within 4 to 6\,ps of simulated time. Of the remaining events, 673 cases (13.5\%) resulted in He thermalization but no Frenkel pairs, further adding to the i-He tally.

The remaining 1,634 cases (32.7\%) deserve special attention. This group includes those He atoms that have thermalized within the core region and have created stable Frenkel pairs but have not become 
substitutional during the MD simulation. The number of stable point defects created by the He collisions as a function of PKA energy is shown in \fref{fig:fp}. Although in principle, He can also be trapped by SIAs \cite{ventelon2006}, we did not observe any instances where this occurred. SIAs created during the cascade either diffused away from the core region or were situated too far from the final position of the He atom.

\begin{figure}
  \centering
  \includegraphics[width=0.9\columnwidth]{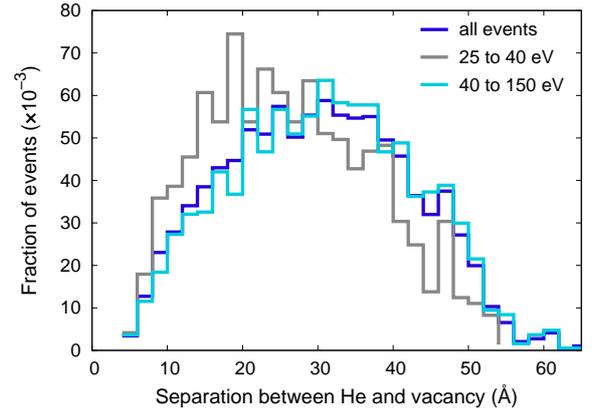}  
  \caption{
    Pair correlation of He interstitials and Fe vacancies obtained from 1,634 cases of He impact in a Fe lattice. Comparison of the relative contributions of different energy windows shows a slight shift toward smaller distances for smaller energies.
  }
  \label{fig:distance}
\end{figure} 

Analysis of the resulting configurations enables us to determine the spatial correlation of He interstitials with vacancies as shown in \fref{fig:distance}. The overall distribution has a mean of approximately 30\,\AA, with lower energy contributions slightly shifted to shorter separations. In any case, to determine the fate of these He ions conclusively, one needs to ``age'' these configurations further using a technique capable of probing longer time scales. This is akin to calculating the \emph{fraction of freely migrating defects} in high-energy cascade simulations \cite{soneda1998}. To this end, we carry out kinetic Monte Carlo (kMC) simulations of He-vacancy reactions according to the distribution given in \fref{fig:distance}.

\subsection{Kinetic Monte Carlo simulations of He-vacancy reactions}
\label{subsec:kmc}

\begin{table*}
  \centering
  \caption{
    Summary of results from MD and kMC simulations of He impacts in BCC Fe. For the data obtained from kMC simulations the third column states the number of occurences equivalent to the number of events treated by MD simulations. The actual number of events simulated by kMC was $10^6$ (see text for details).
  }
  \label{tab:events}
  \begin{tabular}{llcc}
    \hline\hline
    & Event
    & Number of occurences
    & Relative fraction
    \\
    \hline
    i-He
    & He escaped core region during cascade (MD)        &  2668 & 53.3\%\ \\
    & no stable defects created during cascade (MD)     &   673 & 13.5\%\ \\
    & He escaped core region after thermalization (kMC) & ~1513 & 30.3\%\ \\
    & total                                             &       & 97.1\%\ \\[6pt]
    s-He
    & He trapped during cascade (MD)                    &    25 &  0.5\%\ \\
    & He trapped after thermalization (kMC) (MD)        &  ~121 &  2.4\%\ \\
    & total                                             &       &  2.9\%\ \\
    \hline\hline
  \end{tabular}
\end{table*}

The kMC simulations consisted of a single vacancy located at the center of a reaction sphere and a randomly-oriented He atom separated from the vacancy by a distance sampled from the distribution shown in \fref{fig:distance}. We have used the Green's function Monte Carlo method \cite{mhk} in a continuum Fe medium with two spherical particles representing the He atom and the vacancy. The sum of the radii of the vacancy and He atom was set equal to the third-nearest neighbor distance in the BCC lattice, $r=a_0\sqrt{2}$ to be consistent with the He-V binding energy calculations performed in the Appendix. Other authors have suggested that binding occurs up to the fifth nearest neighbor distance \cite{dimitri}. No further correlation between the He atom and the vacancy is assumed, {\it i.e.} jumps toward and away from the central vacancy are sampled with equal probability. The critical parameters for the kMC simulations are the temperature, the diffusivities, and the size of the simulation box. The temperature was 700\,K, the same as in the MD simulations. With respect to the diffusivities, we neglect vacancy diffusion, as its diffusion coefficient is known to be several orders of magnitude lower than that of He, even at these temperatures. The diffusion coefficient of He in BCC Fe was taken from Terentyev {\it et al.} \cite{dimitri}, who used the same interatomic potential as in the present work and obtained $D_\He=5.1\times10^{-3}\exp{\left(-76/kT\right)}$ cm$^2$s$^{-1}$ (migration energy in meV). 
Finally, also for consistency with the  MD simulations, we considered a spherical region with radius equal to $18.65a_0$. 

If over the course of a simulation the He atom escaped the spherical region, it was added to the i-He count, whereas, if it reacted with the vacancy in the center of the simulation sphere, it was tallied as a s-He. After $10^6$ events, we calculated the probability for He-vacancy recombination under these conditions to be 7.4\%. The probability of He escape in an infinite medium has a known analytical solution given by \cite{hudson}
\begin{equation}
  p = 1-\frac{2r}{d}
\end{equation}
where $d$ is the initial separation distance. This formula yields an average reaction probability of approximately 12\%, which is higher than the kMC value because it is not limited to a finite reaction volume.

When prorated to the number of cases that constitute the He-vacancy pair correlation in \fref{fig:distance}, this result, added to the 0.5\% computed directly from the MD simulations, results in a total fraction of s-He of 2.9\%\ in ion implanted BCC Fe. The various contributions to this number are summarized in \tref{tab:events} for clarity.

\section{Discussion and conclusions}
\label{sec:disc}

Understanding how implanted or transmutation He is partitioned between i-He and s-He is paramount because of the different characteristics of both species in the BCC lattice. A comprehensive review on the role of He in Fe has been recently published \cite{samaras2009}, where the basic energetics are given and a review of the existing literature is provided. All studies agree that s-He is energetically more favorable with a lower formation energy and a strong He-vacancy binding energy but cannot move unless aided by other point defects. For its part, i-He diffuses very fast in all three dimensions, rapidly probing vast regions of configuration space and finding sinks or other defects very efficiently. These different behaviors have important implications in terms of the long term microstructural evolution. For example, He in solution in the BCC lattice is known to stabilize vacancy clusters produced directly in high-energy cascades. In terms of its effect on direct damage production, however, the evidence reported in the literature is contradicting. On the one hand, some researchers using the Fe--He Wilson-Johnson potential \cite{wilson} have found that high-energy cascades in BCC Fe doped with small concentrations of s-He result in higher numbers of vacancy clusters than in pure Fe \cite{yang2006}. They also found these clusters to be generally larger in size. On the other hand, using the Juslin-Nordlund potential ---employed here--- Lucas and Sch\"aublin have found that it is i-He in solution that causes larger cluster sizes and number densities to appear \cite{lucas2008}. In fact, they observe that s-He reduces the number of stable defects with respect to pure Fe. These workers also performed a systematic Fe--He potential comparison, noting that the Juslin potential gives results that are overall in better agreement with DFT calculations \cite{fu2007}. In any case, these results show the importance of determining the correct partition of implanted He, something typically neglected in most rate theory studies, where He is generally inserted as i-He (although some notable exceptions exist \cite{katoh1994}). 

Next we discuss the validity and limitations of our approach. This work hinges on the fact that Fe recoil spectra from He ions with a wide range of incident energies are almost identical. Further, assuming that recoils produce isolated cascades, data extracted from ion beam experiments can be used to infer the behavior of $\alpha$ particles created in $(n,\alpha)$ reactions. Then, the only difference between $(n,\alpha)$ reactions and He-ion irradiations is that the former are created homogeneously within the matrix, whereas ions penetrate a short distance into the material (c.f. Table \ref{tab:t_ions}), but up to the cooling-down phase of the cascade, damage is produced in an identical fashion in both instances.

With regard to the interatomic potentials used, we have already mentioned the studies in Refs.~\cite{lucas2008,samaras2009}. Stewart {\it et al.} \cite{stewart2010} have recently carried out an exhaustive comparison of Fe--He and He--He potentials available in the literature. These authors noted that the Fe--He potential is strongly dependent on the matrix (Fe--Fe) to which it is coupled, and that the potentials used in the present work produce little clustering. In addition, Yang {\it et al.} \cite{yang2008} and Pu {\it et al.} \cite{pu2008} have shown that different potentials can have a noticeable influence on vacancy cluster formation in displacement cascades. All of these effects should again be mitigated at high temperatures, yet in light of these results some variability in the final fraction of s-He can be expected.

Another, limitation is the assumption of uncorrelated diffusion in the kMC calculations. Presumably, the migration energy of an i-He varies as a function of its proximity to a vacancy from the value in the bulk (here $E_m=58\,\meV$, see \fref{fig:barrs1}) to zero in the two interstitial sites closest to a vacancy (leading to spontaneous recombination). In this work, however, we have neglected this dependency, for two main reasons. First, because at the simulation temperature, the i-He diffusion barriers, which are typically on the order of 40 to 60\,meV (see \tref{tab:mig}), are well below kinetic energy fluctuations, as shown in the analysis provided in the Appendix. Second, this correlated effect is partially captured by adjusting the interaction distance, which was set to 5$^{\mathrm{th}}$ nearest neighbor distance in this work.  Indeed, other (lattice) kMC calculations of He and He-vacancy complex diffusion have also disregarded this effect for similar reasons \cite{wirth2004}.  

In conclusion, we have performed a computational study of He implantation in BCC Fe and have calculated the fraction of implanted He that becomes substitutional during its own collisional phase. We find that a fraction of approximately 3\%\ is reasonable for the energy range of interest to fusion materials. Damage accumulation calculations should take this datum into account, although more calculations are needed to accuratelt quantify the impact on long-term microstructural evolution.

\section*{Acknowledgments}

This work was performed under the auspices of the U.S. Department of Energy by Lawrence Livermore National Laboratory under Contract DE-AC52-07NA27344. We acknowledge support from the Laboratory Directed Research and Development Program under project 09-SI-003 and allocations of computer time at the National Energy Research Scientific Computing Center, which is supported by the Office of Science of the U.S. Department of Energy under Contract No. DE-AC02-05CH11231.

\theendnotes

\appendix

\section{Helium interstitial migration in the vicinity of an iron vacancy}
\label{app}

To supplement the kMC simulations conducated as part of the present work, we carried out a systematic study of He interstitialimigration in the vicinity of a vacancy. Calculations of migration barriers for He interstitials in defect-free iron and in the vicinity of a vacancy were carried out using the drag method implemented by the authors in the MD code \textsc{lammps}.
All barrier calculations were carried out at the zero-K lattice constant of 2.855\,\AA\ using $16\times 16\times 16$ supercells based on the conventional BCC unit cell.

\begin{figure}
  \centering
  \includegraphics[width=0.9\columnwidth]{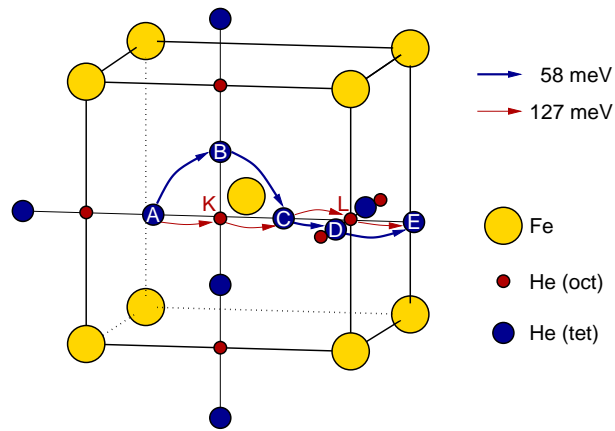}
  \caption{
    Conventional unit cell of the body-centered cubic lattice illustrating the locations of tetrahedral and octahedral interstitial sites, and He migration barriers in the defect-free iron. Note that for clarity only a subset of the interstitial sites is shown.
  }
  \label{fig:barrs1}
\end{figure}

The formation energies of interstitial helium in tetrahedral and octahedral sites are 4.39\,eV and 4.52\,eV, respectively. In the ideal lattice the coordinates of the tetrahedral and octahedral interstitial are $(0, \frac{1}{4}, \frac{1}{2})$ and $(0, 0, \frac{1}{2})$ in units of the lattice constant. As can be deduced from \fref{fig:barrs1}, there are 12 tetrahedral and 6 octahedral sites in the conventional BCC unit cell. \Fref{fig:barrs1} also summarizes the results for the migration barriers of tetrahedral He interstitials in defect-free iron. We obtain a barrier of 58\,meV for migration along $\left<110\right>$ in agreement with Ref.~\cite{juslin2008}. For migration along $\left<100\right>$ the calculations yield a value of 127\,meV, which precisely corresponds to the enegry difference between the tetrahedral and octahedral sites as the latter coincides with the saddle point.


Before we address the migration of He in the vicinity of a vacancy, we first study the effect of the vacancy strain field on the energetics of He interstitials. To this end, we constructed all crystallographically distinct He-vacancy pairs that can occur within a radius of four lattice constants, which yields 78 distinct pairs.

\begin{table}
\centering
\caption{
  Migration barriers for He interstitials in the vicinity of a vacancy.
  $x_{ini}$: initial position in fractional coordinates with respect to the vacancy located at the origin,
  $N_{ini}$: initial neighbor shell,
  $x_{fin}$: final position,
  $N_{fin}$: final neighbor shell,
  $\Delta E_{i-f} = E_f - E_i$: energy difference (meV),
  $\Delta E_b$: migration barrier (meV).
}
\label{tab:mig}
\begin{tabular}{*{4}c*{4}c*{2}r}
\hline\hline
  \multicolumn{3}{c}{$x_{ini}$}
& \multicolumn{1}{c}{$N_{ini}$}
& \multicolumn{3}{c}{$x_{fin}$}
& \multicolumn{1}{c}{$N_{fin}$}
& \multicolumn{1}{c}{$\Delta E_{i-f}$}
& \multicolumn{1}{c}{$\Delta E_b$} \\
\hline

  1             & $\frac{1}{2}$ & $\frac{1}{4}$ & 3       & 1             &  $\frac{3}{4}$ &  $\frac{1}{2}$ &  4     &   $-9$  &  31 \\
                &               &               &         & 1             &  $\frac{1}{4}$ &  $\frac{1}{2}$ &  3     &      0  &  45 \\
                &               &               &         & $\frac{5}{4}$ &  $\frac{1}{2}$ &  0             &  5     &     85  & 123 \\
                &               &               &         & 0             &  0             &  0             &  S     & $-1945$ &  16 \\
                &               &               &         & 1             &  $\frac{1}{2}$ &  $\frac{3}{4}$ &  4     &    $-9$ &  31 \\
                &               &               &         & 0             &  0             &  0             &  S     & $-1945$ &  23 \\

  1             & $\frac{3}{4}$ & $\frac{1}{2}$ & 4       & 1             &  $\frac{1}{2}$ &  $\frac{3}{4}$ &  4     &      0  &  17 \\
                &               &               &         & 1             &  $\frac{1}{2}$ &  $\frac{1}{4}$ &  3     &      9  &  40 \\
                &               &               &         & $\frac{5}{4}$ &  1             &  $\frac{1}{2}$ &  6     &     69  & 120 \\
                &               &               &         & $\frac{3}{4}$ &  1             &  $\frac{1}{2}$ &  4     &      0  &   4 \\
                &               &               &         & 1             &  $\frac{1}{4}$ &  $\frac{1}{2}$ &  3     &      9  &  40 \\
                &               &               &         & 1             &  $\frac{5}{4}$ &  $\frac{1}{2}$ &  6     &     69  & 120 \\

  $\frac{5}{4}$ & $\frac{1}{2}$ & 0             & 5       & $\frac{3}{2}$ &  $\frac{3}{4}$ &  0             &  8     &   $-20$ &  45 \\ 
                &               &               &         & $\frac{3}{2}$ &  $\frac{1}{4}$ &  0             &  7     &     12  &  78 \\
                &               &               &         & 1             &  $\frac{1}{2}$ &  $\frac{1}{4}$ &  3     &   $-85$ &  38 \\
                &               &               &         & 1             &  $\frac{1}{2}$ & -$\frac{1}{4}$ &  3     &   $-85$ &  38 \\
                &               &               &         & $\frac{7}{4}$ &  $\frac{1}{2}$ &  0             & 13     &   $-18$ &  55 \\
                &               &               &         & 0             &  0             &  0             &  S     & $-2030$ &  59 \\

  $\frac{5}{4}$ & 1             & $\frac{1}{2}$ & 6       & $\frac{3}{2}$ &  1             &  $\frac{3}{4}$ & 10     &      2  &  63 \\
                &               &               &         & $\frac{3}{2}$ &  1             &  $\frac{1}{4}$ &  9     &      1  &  56 \\
                &               &               &         & 1             &  $\frac{5}{4}$ &  $\frac{1}{2}$ &  6     &      0  &  56 \\
                &               &               &         & 1             &  $\frac{3}{4}$ &  $\frac{1}{2}$ &  4     &   $-69$ &  51 \\
                &               &               &         & $\frac{7}{4}$ &  1             &  $\frac{1}{2}$ & 14     &      7  &  58 \\
                &               &               &         & $\frac{3}{4}$ &  1             &  $\frac{1}{2}$ &  4     &   $-69$ &  51 \\

  $\frac{3}{2}$ & $\frac{1}{4}$ & 0             & 7       & $\frac{7}{4}$ &  $\frac{1}{2}$ &  0             & 13     &   $-30$ &  35 \\
                &               &               &         & $\frac{5}{4}$ &  $\frac{1}{2}$ &  0             &  5     &   $-12$ &  66 \\
                &               &               &         & $\frac{3}{2}$ &  0             &  $\frac{1}{4}$ &  7     &      0  &  58 \\
                &               &               &         & $\frac{3}{2}$ &  0             & -$\frac{1}{4}$ &  7     &      0  &  58 \\
                &               &               &         & $\frac{3}{2}$ &  $\frac{3}{4}$ &  0             &  8     &   $-32$ &  40 \\
                &               &               &         & $\frac{3}{2}$ & -$\frac{1}{4}$ &  0             &  7     &      0  & 125 \\

\hline\hline
\end{tabular}
\end{table}

The binding energy for a He-vacancy pair is defined as
\begin{equation}
  \Delta E_b = \Delta E_f(\mathrm{He}-\mathrm{V}) - \Delta E_f(\mathrm{He}) - \Delta E_f(\mathrm{V}),
\end{equation}
where $\Delta E_f(\mathrm{He}-\mathrm{V})$, $\Delta E_f(\mathrm{He})=4.391\,\eV$, and $\Delta E_f(\mathrm{V})=1.721\,\eV$ are the formation energies of the He-vacancy pair, the isolated He interstitial as well as the isolated vacancy. Negative and positive values of $\Delta E_b$ indicate attraction and repulsion of the He-vacancy pair, respectively.  The formation energy is given by
\begin{equation}
  \Delta E_f = E_\mathrm{def} - \frac{N_\mathrm{def}}{N_\mathrm{id}} E_\mathrm{id},
\end{equation}
where $E_i$ and $N_i$ are the total energy and the number of Fe atoms in configuration $i$. Here, we have quietly set the chemical potential of He to zero which has a minimal effect on the formation energies and no effect on the binding energy.

\begin{figure}
  \centering
  \includegraphics[width=0.9\columnwidth]{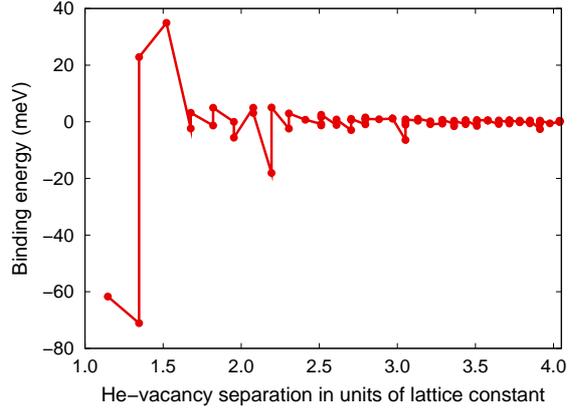}
  \caption{
    Binding energy of He-vacancy pairs as a function of separation after relaxation.
  }
  \label{fig:bind}
\end{figure}

The binding energy is shown as a function of the He-vacancy separation in \fref{fig:bind}. Note that the two nearest He-vacancy pairs
have been omitted in \fref{fig:bind}, since they spontaneously recombine with the vacancy leading to a strongly negative binding energy of $-2.01$ eV. We find notable variations in the binding energy at least up to a separation of about 2.2 lattice constants. It is furthermore noteworthy that the variation in the binding enegry is not monotonic with the distance but exhibits pronounced oscillations that to some extent can be correlated with different crystallographic directions.

We can now consider the different possibilities for He interstitials migrating in the vicinity of a vacancy. From \fref{fig:barrs1}, we can deduce that from any tetrahedral site there are six possible jumps, four of which are along $\left<110\right>$ and two of which are along $\left<100\right>$. More specifically,
\begin{enumerate}
\item
  tet $\rightarrow$ tet:
  $(0, \frac{1}{4}, \frac{1}{2}) + (0, +\frac{1}{4}, +\frac{1}{4})
  \rightarrow (0, \frac{1}{2}, \frac{3}{4})$
\item
  tet $\rightarrow$ tet:
  $(0, \frac{1}{4}, \frac{1}{2}) + (0, +\frac{1}{4}, -\frac{1}{4})
  \rightarrow (0, \frac{1}{2}, \frac{1}{4})$
\item
  tet $\rightarrow$ tet:
  $(0, \frac{1}{4}, \frac{1}{2}) + (+\frac{1}{4}, -\frac{1}{4}, 0)
  \rightarrow (\frac{1}{4}, 0, \frac{1}{2})$
\item
  tet $\rightarrow$ tet:
  $(0, \frac{1}{4}, \frac{1}{2}) + (-\frac{1}{4}, -\frac{1}{4}, 0)
  \rightarrow (-\frac{1}{4}, 0, \frac{1}{2})$
\item
  via oct:
  $(0, \frac{1}{4}, \frac{1}{2}) + (0, +\frac{1}{2}, 0)
  \rightarrow (0, \frac{3}{4}, \frac{1}{2})$
\item
  via oct:
  $(0, \frac{1}{4}, \frac{1}{2}) + (0, -\frac{1}{2}, 0)
  \rightarrow (0, -\frac{1}{4}, \frac{1}{2})$
\end{enumerate}

The barriers that are obtained for these paths starting from different initial He-vacancy arrangements are summarized in Table \ref{tab:mig}. We find that while there is a great variability of the values, in general migration barriers range from about 20 to 50\,meV for shells 3 and 4, and from 40 to 60\,meV for shells 5 to 7, relatively quickly closing in on the bulk value of 58\,meV. Using these data, we rationalized (see \sref{sec:disc}) that a good choice for the interaction radius entering the kMC simulations is five lattice constants.

\bibliographystyle{elsarticle-num}

\end{document}